\documentclass{article}
\usepackage{spconf, amsmath, amssymb, bm, graphicx, url}
\usepackage[dvipsnames]{xcolor}

\usepackage[some,bottom]{background}

\SetBgContents{
\begin{minipage}{10in}
\begin{center}
{I. Jang, H. Yang, W. Lim, S. Beack and M. Kim, ``Personalized Neural Speech Codec,"}\\ 
{IEEE International Conference on Acoustics, Speech and Signal Processing (ICASSP), Seoul, Korea, 2024, pp. 991-995, doi: 10.1109/ICASSP48485.2024.10446067.}        
\end{center}
\end{minipage}
}
\SetBgScale{.8}
\SetBgOpacity{1}
\SetBgColor{black}
\SetBgVshift{1cm}

\title{Improving Neural Speech Compression through Personalization}
\title{Personalized Neural Speech Codec}
\name{Inseon Jang$^1$, Haici Yang$^2$, Wootaek Lim$^1$, Seungkwon Beack$^1$, and Minje Kim$^3$\thanks{This work was supported in part by Electronics and Telecommunications Research Institute (ETRI) grant funded by the Korean government [23ZH1200: The research of the basic media contents technologies] as well as by the National Science Foundation under Grant No. 2046963.}
\sthanks{Work done at Indiana University.}}
\address{$^1$Electronics and Telecommunications Research Institute, Daejeon, South Korea 34129\\
$^2$Indiana University, Department of Intelligent Systems Engineering, Bloomington, IN, USA 47408\\
$^3$University of Illinois at Urbana-Champaign, Department of Computer Science, IL, USA 61801}
\begin{document}
\ninept
\maketitle
\begin{abstract}
In this paper, we propose a personalized neural speech codec, envisioning that personalization can reduce the model complexity or improve perceptual speech quality. Despite the common usage of speech codecs where only a single talker is involved on each side of the communication, personalizing a codec for the specific user has rarely been explored in the literature. First, we assume speakers can be grouped into smaller subsets based on their perceptual similarity. Then, we also postulate that a group-specific codec can focus on the group's speech characteristics to improve its perceptual quality and computational efficiency. To this end, we first develop a Siamese network that learns the speaker embeddings from the LibriSpeech dataset, which are then grouped into underlying speaker clusters. Finally, we retrain the LPCNet-based speech codec baselines on each of the speaker clusters. Subjective listening tests show that the proposed personalization scheme introduces model compression while maintaining speech quality. In other words, with the same model complexity, personalized codecs produce better speech quality. 
\end{abstract}
\begin{keywords}
Speech coding, neural speech coding, personalization, model compression
\end{keywords}

\BgThispage

\section{Introduction}
\label{sec:intro}

The recent advances in neural speech coding (NSC) technology have achieved unprecedented coding gain, which relied significantly on the decoder's generalization power. In representative NSC approaches, a neural vocoder restores the original speech waveforms from their compressed bitstream, often using a generative model. For example, autoregressive models, such as WaveNet \cite{OordA2016wavenet}, have shown transformative coding gain for very low-bitrate speech coding (2.4 kbps \cite{KleijnW2018wavenet} and 1.6 kbps \cite{GarbaceaC2019vqvae}, respectively) thanks to the WaveNet's advanced architecture as well as its the large model capacity (of about 20M parameters). Indeed, the high-quality sample-by-sample prediction of waveform signals comes at the cost of expensive inference complexity, about 100G floating-point operations per second (FLOPS), prohibiting their use in resource-constrained devices. 

There have been recent efforts to improve the efficiency of NSC models. 
LPCNet \cite{ValinJ2019lpcnet} successfully harmonized the linear predictive coding (LPC) and WaveRNN \cite{KalchbrennerN2018wavernn} by training the WaveRNN module to predict the excitation of the speech, i.e., the residual of the linear prediction, rather than the raw audio samples. Consequently, the complexity of an LPCNet-based decoder is as low as around 3 GFLOPS with 30 MFLOPS for encoding \cite{ValinJ2019lpcnetcoding, Subramani2022e2elpcnet} or lower \cite{Valin2022lpcnetimproving}. Their low arithmetic and spatial complexity, however, come at the cost of suboptimal sound quality compared to the WaveNet decoder.

Autoencoders are another choice for a low-complexity NSC, where the system consists of a pair of encoder and decoder modules. With the encoder's capability of learning the compact code representation, the decoder can be streamlined accordingly. Soundstream employed residual vector quantization and fully-convolutional encoder and decoder, outperforming traditional speech codecs \cite{Zeghidour2021soundstream}. Although its inference runs in real-time on a single smartphone CPU, it still requires 11 GFLOPS to decode. EnCodec, based on the SoundStream model, improved sound quality further with an additional adversarial loss \cite{defossez2022highfi}. It adopted a lightweight Transformer \cite{VaswaniA2017transformer} for additional coding gain but at the cost of increased algorithmic delay. Recently reported NSCs introduced various advantages in low bitrates, such as T-F codec \cite{JiangX2023tf-codec} for low latency, DAC \cite{kumar2023high} and PostGAN \cite{KorseS2022postgan} for high sound quality, etc., but they rarely targeted the efficiency goal.
Likewise, there is a tradeoff between the model complexity and coding gain in the NSC literature, which we tackle in this paper by proposing \textit{personalized} neural speech coding (PNSC). 


Personalization has shown promising results in model compression tasks for speech enhancement \cite{SivaramanA2022jstsp, SivaramanA2021waspaa, KimSW2021waspaa, Thakker2022arxiv}. A personalized model adapts to the target speaker group's speech trait, narrowing the training task down to a smaller subtask, i.e., defined by the smaller speaker group than the entire speakers in the corpus. As a result, the personalized model can be seen as a more compact and specialized version of the computationally complex generalist model. 


In legacy speech and audio codecs, adaptive coding is a commonly used concept. MPEG-D Unified Speech and Audio Coding (USAC) \cite{usac1} and 3GPP Enhanced Voice Services (EVS) \cite{evs} selectively use coding modules by classifying the signal characteristics. Since each module is specialized in the specificity of the given audio frame (e.g., whether it contains transient or not), these hybrid systems outperform their predecessors. PNSC is based on similar principles to hybrid codecs': we postulate that there exists a specialized module more suitable than the others for the given input signal's characteristics (i.e., the test speaker's speech trait). 

In this paper, we focus on the model compression aspect of personalization, while we assume a single user at each end of the speech communication. In addition, we aim to improve the perceptual quality of the baseline when it is compared to the PNSC model with the same bitrate and decoder complexity.
To this end, we begin from LPCNet \cite{ValinJ2019lpcnet}, which has been proven to be one of the most compact decoding schemes in the context of NSC \cite{ValinJ2019lpcnetcoding}. We show that personalization can introduce an additional complexity reduction to this tight LPCNet baseline with an improved adaptation to the talker's personal speech characteristics. In particular, we perform clustering on the speaker embeddings learned from the LibriSpeech \cite{PanayotovV2015Librispeech} dataset via Siamese network-based contrastive learning \cite{BromleyJ1994nips}. Then, we train the LPCNet decoders, each of which is dedicated to recovering the corresponding cluster's utterances. Since we assume the talker's identity does not change frequently, the system operates with minimal additional overhead. In addition, the speaker-specific decoders are trained with potential misclassification errors, making them robust to real-world use cases. 

\begin{figure}
    \centering
    \includegraphics[width=0.85\columnwidth]{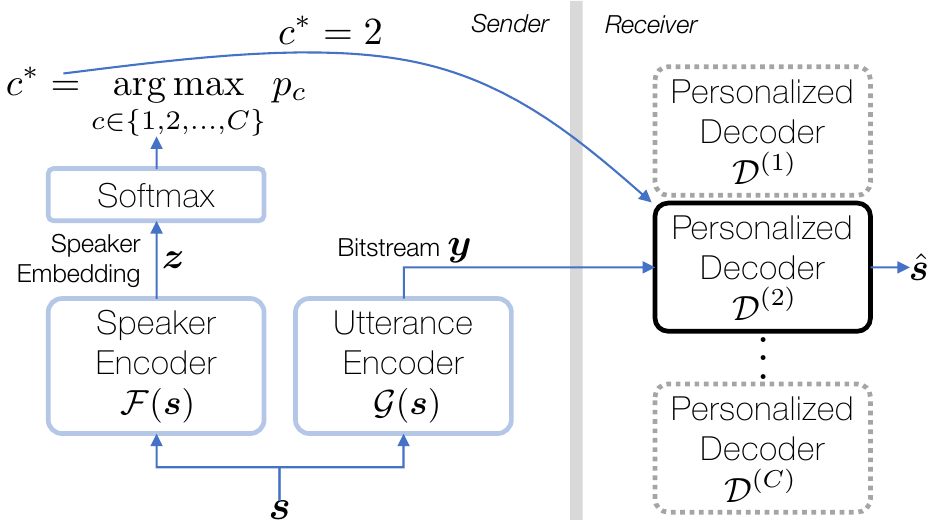}
    \vspace{-0.1in}
    \caption{The overview of the personalized NSC system used in the speech communication scenario. Note that the utterance encoder model $\mathcal{G}(\cdot)$ could be also specialized to handle a specific speaker group, while we leave that option to future work.}
    \label{fig:overview}
\end{figure}

\section{Personalized Neural Speech Codec}
\label{sec:proposed}
We propose personalized LPCNet (PLPCNet), which consists of multiple speaker group-specific LPCNet decoders. From the sender side, the bitstream (i.e., quantized speech features) is transmitted to the receiver, along with speaker group index if necessary. Then, the corresponding PLPCNet decoder is chosen on the receiver side to reconstruct the speech utterance. In this section, we first describe how to define speaker groups from the large speech corpus. Then, PLPCNet is defined as a specialist version of the original LPCNet. 


\subsection{The Overview of the System for Speech Communication}
\label{ssec:overview}

Fig. \ref{fig:overview} illustrates the general PNSC concept used in the speech communication scenario, where the receiver knows which speaker group the sender belongs to to choose the right personalized decoder. Hence, a speaker classification decision must precede on the sender side and be delivered to the receiver in the form of the group index. 

\noindent\textbf{Speaker Classification}: Let our \textit{speaker encoder} $\mathcal{F}(\cdot)$ be a function that converts an input speech signal $\bm{s}$ into a feature vector $\bm{z}$ that contains speaker-specific discriminative information: $\bm{z}\leftarrow\mathcal{F}(\bm{s}), ~\bm{z}\in\mathbb{R}^D$. 
Then, the $D$-dimensional speech feature vector $\bm{z}$ goes through a softmax layer to estimate the posterior probability vector $\bm{p}\in\mathbb{R}^C$ over $C$ total speaker groups: $[p_1,p_2,\ldots,p_C]\!\leftarrow\!\text{softmax}(\bm{z})$.
Finally, the best group $c^*=\operatornamewithlimits{\arg \max}_{c\in\{1,2,\ldots,C\}} p_c$ is selected. 

\noindent\textbf{Encoding}: Meanwhile, the input utterance $\bm{s}$ is also fed to a separate \textit{utterance encoder} $\mathcal{G}(\cdot)$ to acquire a compact bitstring $\bm{y}\leftarrow\mathcal{G}(\bm{s}), ~\bm{y}\in\{0,1\}^L$. Although the PNSC concept extends to personalizing the generic encoder model $\mathcal{G}(\cdot)$ into the $c$-th group, i.e., $\mathcal{G}^{(c)}(\cdot)$, in this paper, we inherit the LPCNet's simple cepstrum-based code produced from a deterministic encoding function $\mathcal{G}(\cdot)$. 

\noindent\textbf{Decoding}: Instead, we focus on personalizing the LPCNet-based decoder. With the received speech code $\bm{y}$ and the speaker group index $c^*$, the corresponding decoder is chosen to recover the original speech $\bm{s}$ from the code $\bm{y}\approx\hat{\bm{s}}\leftarrow\mathcal{D}^{(c^*)}(\bm{y})$,
where $\mathcal{D}^{(c)}(\cdot)$ denotes the $c$-th decoder prepared ahead of time to handle the $c$-th speaker group. Note that the bitstream $\bm{y}$, i.e., the quantized code, is not necessarily the same as $\bm{z}$ because the speaker embedding is learned to discriminate different speakers instead of conveying information for recovering perceptually meaningful speech signals. 

\subsection{The Siamese Network for Speaker Embedding Learning}
\label{ssec:embedding}


PNSC assumes that there is a semantically cohesive group of speakers, who share similar speech characteristics. Hence, it is also assumed that the variation within the subset is lower than the entire speech corpus. Typically, the training objective of an ordinary deep learning model is to generalize to the large data variation, requiring a large model capacity. Conversely, we focus on a cohesive subset of the data, where smaller models suffice. 

Likewise, a reasonable grouping strategy of speakers is key to successful personalization. Following \cite{SivaramanA2021waspaa}, we first learn the speaker encoder $\mathcal{F}(\cdot)$ that learns discriminative speaker embeddings $\bm{z}$. They define the embedding space, where clustering is performed. 

We employ the Siamese network principle \cite{BromleyJ1994nips, ChiccoD2021siamese} to train $\mathcal{F}(\cdot)$ in a contrastive way using positive and negative pairs of speech utterances. From a given set of utterances spoken by the $k$-th speaker $\mathbb{S}^{(k)}$, we sample two utterances $\bm{s}_i$ and $\bm{s}_j$, i.e., $i,j\in\mathbb{S}^{(k)}$, which the Siamese network encoder takes as input and performs inference on each of them, respectively: $\bm{z}_i\leftarrow\mathcal{F}(\bm{s}_i), ~~\bm{z}_j\leftarrow\mathcal{F}(\bm{s}_j)$. Then, $\bm{z}_i$ and $\bm{z}_j$, are compared to improve their similarity, as they originate from the same speaker. Meanwhile, the negative pair is also sampled, but from two different speakers, $\mathbb{S}^{(k)}$ and $\mathbb{S}^{(k')}$, respectively, whose corresponding embeddings are supposed to be different from each other. We represent this process as a binary cross-entropy loss:
\begin{equation}
    \mathcal{L}_\text{emb}=-\!\!\!\!\!\!\sum_{\substack{i,j\sim\mathbb{S}^{(k)},\\~\forall k}}\!\!\!\log \sigma(\bm{z}_i^\top\bm{z}_j)
    -\!\!\!\!\!\!\!\!\!\sum_{\substack{i\sim\mathbb{S}^{(k)}, j\sim\mathbb{S}^{(k')},\\ ~k\neq k'}}\!\!\!\!\!\!\!\!\! \log \big(1- \sigma(\bm{z}_i^\top\bm{z}_j)\big),
\end{equation}
where the inner product between the two embeddings is used to measure the level of agreement, followed by the sigmoid function $\sigma(\cdot)$ to turn the quantity into probabilistic values.



\subsection{Speaker Clustering}

To determine the $C$ speaker groups, we perform k-means clustering in the embedding space defined by the speaker encoder $\mathcal{F}(\cdot)$, assuming that the intrinsic nonlinear relationship between speakers is represented in the Euclidean space linearly. In particular, the Siamese network $\mathcal{F}(\cdot)$ is learned to represent such linear similarity (i.e., inner products) in the latent space. 

Clustering is done on the speaker embedding $\bar{\bm{z}}^{(k)}$, which is the average of all utterance-specific embeddings that belong to the $k$-th speaker: $\bar{\bm{z}}^{(k)}=\frac{1}{|\mathbb{S}^{(k)}|}\sum_{i\in\mathbb{S}^{(k)}}\bm{z}_i$, where $|\mathbb{S}^{(k)}|$ denotes the number of elements in the set. The resulting $C$ centroids are represented by $[\bm{h}^{(1)}, \bm{h}^{(2)}, \ldots, \bm{h}^{(C)}]$, each of which is the average of the speaker embeddings that belong to the corresponding cluster, i.e., $\bm{h}^{(c)}=\frac{1}{|\mathbb{H}^{(c)}|}\sum_{k\in\mathbb{H}^{(c)}}\bar{\bm{z}}^{(k)}$, where $\mathbb{H}^{(c)}$ denotes the $c$-th speaker cluster.

The offline k-means clustering process groups the speakers in the training set into $C$ classes, which will serve as the ground-truth class labels for the speaker classification task. A simple softmax layer can convert the embedding $\bm{z}$ into a probability vector as described in Sec. \ref{ssec:overview}. In practice, we opt to perform classification by finding the nearest cluster centroid of the given speaker embedding, skipping the explicit use of the softmax layer. 

During the test time, the classification result, $c^*$, defines the speaker group that the receiver has to be based on. Transmission of $c^*$ takes only $\lceil \log_2 C \rceil$ bits, e.g., 2 bits when $C=4$. If we assume that the sender's identity does not change frequently, sending this kind of flag every now and then (e.g., every second) is ignorable. In our experiments, we assume a single-talker scenario, hence the speaker classification happens only once. We leave more dynamic multi-talker scenario to future work.



\subsection{Personalized LPCNet}
\label{ssec:plpcnet}
The decoder consists of a set $\{\mathcal{D}^{(1)},\mathcal{D}^{(2)},\cdots,\mathcal{D}^{(C)}\}$, each of which is a specialist version of the generic LPCNet, i.e., PLPCNet. 

\begin{figure}
    \centering
    \includegraphics[width=0.7\columnwidth]{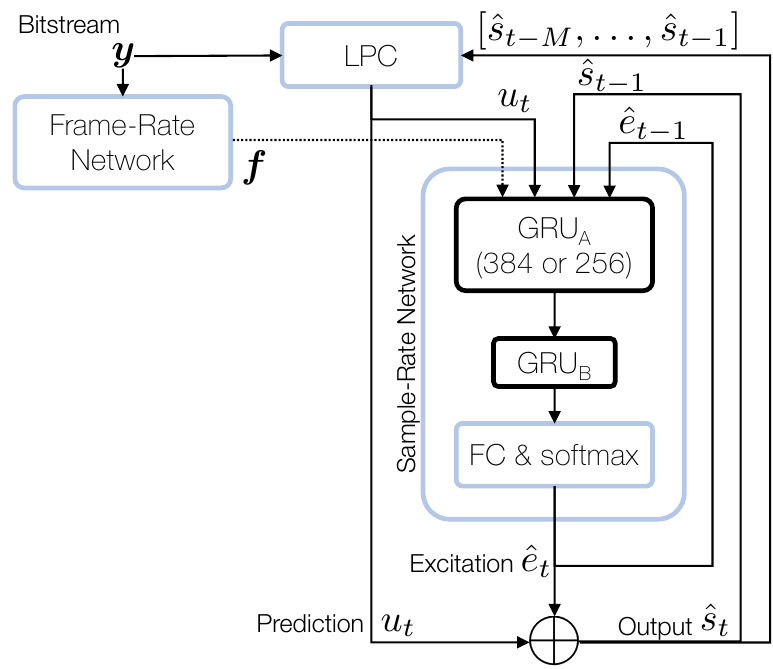}
    \vspace{-0.1in}
    \caption{A simplified LPCNet architecture. PLPCNet controls its complexity by adjusting the number of the GRU layer's hidden units.}
    \label{fig:plpcnet}
\end{figure}

LPCNet is a neural vocoder operating at 1.6 kbps for wide-band speech coding. Its low bitrate is achieved by building on WaveRNN \cite{KalchbrennerN2018wavernn} and combining it with LPC to offload the DNN's reconstruction task. Fig. \ref{fig:plpcnet} illustrates the simplified LPCNet architecture. It is comprised of three main modules: the deterministic LPC module and frame- and sample-rate networks.
In the LPC module, the prediction at time $t$ is obtained from the linear combination of previous speech samples $[s_{t-M}, \ldots, s_{t-1}]$ and their coefficients $[a_{M}, \ldots, a_{1}]$, i.e., $u_t=\sum_{m=1}^M a_m s_{t-m}$, where $M$ is the prediction order. In LPCNet, the coefficients are calculated from the quantized cepstrum in the bitstream $\bm y$. Finally, the excitation is obtained from $e_t=s_t-u_t$.

Meanwhile, the frame-rate network converts $\bm y$ to the frame-rate feature vector $\bm f$ for every 10 ms frame. In the sample-rate network, the previous speech sample $\hat{s}_{t-1}$, previous excitation $\hat{e}_{t-1}$, and current prediction $u_t$ are concatenated to form an input vector, in a sample-by-sample manner. Note here that the decoder takes its own output $\hat{s}_{t-1}$ and $\hat{e}_{t-1}$ instead of the unknown ground-truth samples ${s}_{t-1}$ and ${e}_{t-1}$. It also receives the conditioning input from the frame-rate network, $\bm f$, at every frame. The sample-rate network predicts the probability over the excitation sample, $P(e_t)$, from which the output excitation value  $\hat{e}_t$ is sampled. Finally, adding it to the current LPC prediction $u_t$ generates the current speech sample $\hat{s}_t$. Note that samples are represented in the 8-bit $\mu$-law domain. 

The proposed PLPCNet is trained in a group-specific manner. For the $c$-th speaker group, the estimated excitation of the $i$-th frame bitsting $\bm y_i$ is compared to the ground-truth excitation computed directly from the offline LPC module:
\begin{equation}\label{eq:loss}
    \sum_{i\in\mathbb{S}^{(k)}, k\in\mathbb{H}^{(c)}} \mathcal{L}_\text{CE}(\hat{\bm e}_i||{\bm e}_i), \quad\hat{\bm e}_i\leftarrow\mathcal{D}^{(c)}(\bm y_i),
\end{equation}
where the cross entropy loss $\mathcal{L}_\text{CE}(\cdot)$ between the residual samples is computed in the $\mu$-law space, and then summarized over all training utterances in class $c$. The LPC prediction $\bm u$ is ignored in the loss.

Since LPCNet predicts the LPC residual, the benefit of personalizing it could be limited, compared to a model that works in the raw signal domain. While it makes LPCNet a more challenging baseline to compete with, we still expect that the LPCNet's extended definition of input, i.e., concatenation of the raw samples and the frame-rate feature, could be a suitable representation for personalization.


\begin{figure}[t]
\begin{minipage}[b]{.48\linewidth}
  \centering
  \centerline{\includegraphics[width=3.8cm]{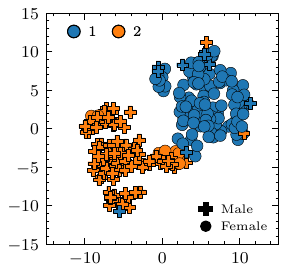}}
  \centerline{(a) Speaker groups for $C=2$}\medskip
\end{minipage}
\hfill
\begin{minipage}[b]{0.48\linewidth}
  \centering
  \centerline{\includegraphics[width=3.8cm]{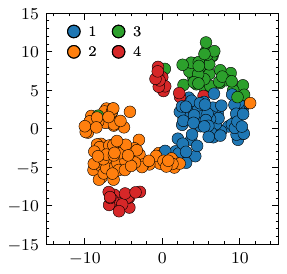}}
  \centerline{(b) Speaker groups for $C=4$}\medskip
\end{minipage}
\vspace{-0.15in}
\caption{Clustering of speakers from different choices of $C$.}
\label{fig:clustering}
\end{figure}

\section{Experiments}
\subsection{Experimental Setup}
\label{ssec:setup}

We use \texttt{train-clean-100} and \texttt{dev-clean} folds from Librispeech \cite{PanayotovV2015Librispeech} whose 16-bit amplitudes were sampled at 16kHz rate. The total length of utterances in \texttt{train-clean-100} is 100 hours, spoken by 251 speakers. Out of them, we set aside 20 speakers for validation. \texttt{dev-clean}'s 40 speakers are used for testing.

The Siamese network $\mathcal{F}(\cdot)$ employs two 32-unit GRU layers as proposed in \cite{SivaramanA2021waspaa}, while we train it without any noise injection. Fig. \ref{fig:clustering} shows different clustering results by varying the number of groups $C$. Each of the 231 points represents one of the $K=231$ training speakers. To visualize them, the original speaker embeddings of $D=32$ are reduced to a 2D space using t-SNE \cite{VanDerMaatenL2008tsne} with the perplexity parameter set to be 40. The subplots show that learned embedding space provides perceptually meaningful discrimination of speakers, e.g., when $C$ = 2 the clusters are formed by the gender of the speakers. In theory, more clusters could lead to better specialization, while the performance gain achieved by personalization saturates at some point, too, as shown in \cite{SivaramanA2021waspaa}. Considering the number of speakers and amount of utterances in each speaker group, we choose to partition the training set into four groups, i.e., $C=4$. As a result, the number of training speakers in cluster is 70, 89, 41, and 31, respectively. Accordingly, the validation set consists of four subsets of 4, 4, 7, and 5 speakers, respectively.

Using LPCNet's open-sourced framework\footnote{\label{fn:lpcnet}https://github.com/xiph/LPCNet}, we extract the bitstream from the Librispeech utterances, i.e., the quantized 18 Bark-scale cepstral coefficients and 2 pitch parameters. 

As baselines, we employ two different versions of the generic LPCNet: large and small architectures. To this end, we build our own PyTorch implementations of LPCNet and train them with the entire Librispeech training fold of 231 speakers. Since the sample-rate network's first GRU layer is the biggest contributor to the computational complexity, we vary its number of hidden units from 384 to 256, which correspond to our large (\texttt{LPCNet-BL-L}) and small (\texttt{LPCNet-BL-S}) models, respectively. In this way, the model size reduces from 1.234M total parameters to 0.784M, which is a 36.47\% reduction. For a fair comparison, we also run the public LPCNet\textsuperscript{\ref{fn:lpcnet}} version with 384 GRU hidden units (\texttt{LPCNet-Pub-L}) on the test sequences and include the results in our listening tests.

Throughout the experiment, we opt for a batch size of 64 and use the Adam optimizer with $\beta_1$ = 0.9, $\beta_2$ = 0.999, and a learning rate of $10^{-3}$ \cite{KingmaD2015adam}. To train PLPCNet models and our own baselines, we also employ the gradient clipping operation to stabilize the RNN training, whose threshold is set to be from $5\times10^{-2}$ to $1\times 10^{-4}$. The smaller the model size and the more speakers are trained, the smaller the threshold is used.

In addition, we train $C=4$ PLPCNet decoders as described in Sec. \ref{ssec:plpcnet}, but by varying the model sizes once again, resulting in eight PLPCNet models in total (four per architecture). We denote them by \texttt{PLPCNet-L} and \texttt{PLPCNet-S}, respectively. 

The subjective listening tests evaluate the perceptual quality of the proposed PLPCNet models. Eight gender- and group-balanced speakers are randomly chosen from \texttt{dev-clean}. The MUSHRA-style \cite{mushra} test contains a hidden reference, a low pass-filtered anchor at 3.5kHz, and the five systems in comparison: \texttt{LPCNet-BL-L}, \texttt{LPCNet-BL-S}, \texttt{LPCNet-Pub-L}, \texttt{PLPCNet-L}, and \texttt{PLPCNet-S}. When the test signal is processed by \texttt{PLPCNet-L} or \texttt{PLPCNet-S}, we use the estimated class label $c^*$ to choose the corresponding group-specific decoder, which is a process that properly simulates the real-world use case. Nine audio and speech experts participated in the listening test, who all passed the screening process.

\subsection{Experimental Results}
\label{ssec:result}

\begin{figure}[t]
\centering
\includegraphics[width=.8\columnwidth]{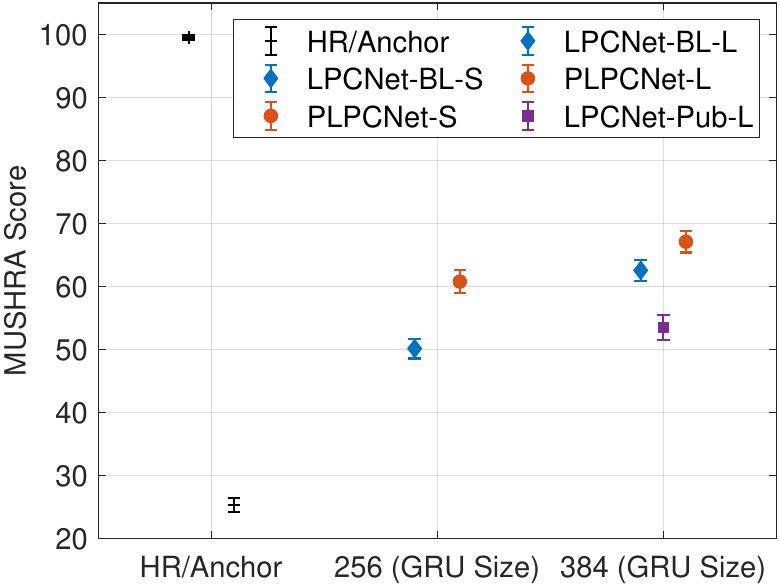}
\vspace{-0.1in}
\caption{Results of the listening test. 95\% confidence intervals are shown as upper and lower bars.}
\vspace{-0.1in}
\label{fig:mushra}   
\end{figure}

Fig. \ref{fig:mushra} shows the MUSHRA-like listening test results. First of all, we observe that our own PyTorch baseline \texttt{LPCNet-BL-L} shows superior performance to the public LPCNet implementation \texttt{LPCNet-Public-L}, while \texttt{LPCNet-BL-S} catches up. We believe that it is due to the different training hyperparameters we tried, including the gradient clipping option. In addition, it is also possible that our models could have been optimized for the Librispeech corpus, which the public LPCNet model might have to generalize to. However, we opt to provide the results from both our own baselines and the public model to note that we are comparing to better, thus more challenging baselines of our own.

The main claims we make in Fig. \ref{fig:mushra} are as follows. First, we see that the proposed PLPCNet models significantly outperform their corresponding (i.e., same-sized) baseline models. This means that the proposed personalization approach introduces an additional performance gain with no cost of increased model complexity or bitrate. We argue that it must be mainly due to the LPCNet decoder's specialization in the speaker group that it is dedicated to. Second, we also see that the \texttt{PLPCNet-S}'s performance catches up \texttt{LPCNet-BL-L}'s and their confidence intervals overlap. These results showcase a model compression ratio of 36.47\% with insignificant performance degradation in terms of sound quality. Third, when we compare the two model sizes, more significant performance improvement is observed when smaller models are in comparison (\texttt{PLPCNet-S} vs. \texttt{LPCNet-BL-S}) than the larger models ((\texttt{PLPCNet-L} vs. \texttt{LPCNet-BL-L}). This trend aligns well with the personalized speech enhancement literature: model personalization benefits compressed model architectures more than the larger ones \cite{SivaramanA2022jstsp, SivaramanA2021waspaa, KimSW2021waspaa}. Finally, it is also worth noting that each test sequence is handled by a selected personalized decoder, where the choice is based on the estimated speaker class. Hence, the listening test results encompass the potential misclassification cases, too.  

\begin{figure}[t]
  \centering
  \centerline{\includegraphics[width=.8\columnwidth]{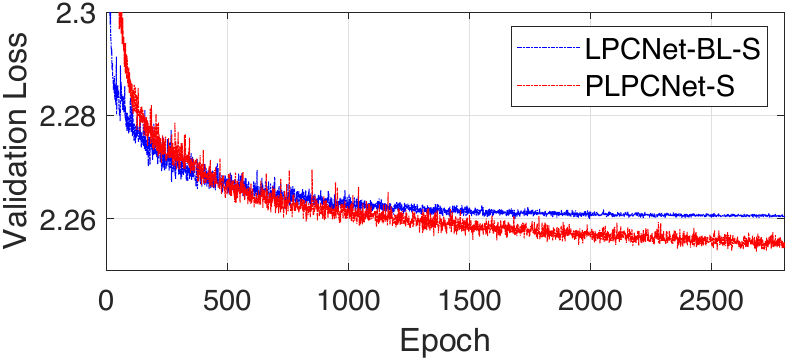}}
  \vspace{-0.1in}
\caption{Comparison of the validation loss curves.}
\vspace{-0.1in}
\label{fig:val_loss}
\end{figure}

In addition, Fig. \ref{fig:val_loss} juxtaposes the validation loss curves of \texttt{PLPCNet-S} and \texttt{LPCNet-BL-S}, where the GRU layer is with 256 hidden units. The PLPCNet graph is a weighted average of validation losses, collected from all $C=4$ decoders as follows:
\begin{equation}
\mathcal{L}_\text{val}=\cfrac{1}{N}\sum_{c=1}^C |\mathbb{H}^{(c)}_\text{val}| \mathcal{L}_\text{val}^{(c)},   \quad N=\sum_{c=1}^C|\mathbb{H}^{(c)}_\text{val}|,
\end{equation}
where $|\mathbb{H}^{(c)}_\text{val}|$ denotes the number of speakers in the $c$-th speaker group in the validation set. The total number of validation speakers $N=20$ in our experiments. $\mathcal{L}_\text{val}^{(c)}$ denotes the average validation loss of the $c$-th speaker group, defined in eq. \eqref{eq:loss}. The weighted sum correctly accounts for the different contributions of the group-specific validation losses depending on the size of the group. Given this, Fig. \ref{fig:val_loss} clearly shows that PLPCNet reaches lower loss values than the corresponding LPCNet baseline, which might have led to its improved subjective quality on the test signals as well. 


\section{Conclusion}
\label{sec:conclusion}
In this paper, we proposed the personalized LPCNet as a promising solution to compressing the NSC models.
We verified the concept in the communication scenario where one specific person was involved on the sender side. We pre-defined four semantically meaningful speaker groups by using discriminative speaker embeddings, and then trained four LPCNet decoders from them, respectively. During the test time, the optimal decoder was estimated to produce the best reconstruction. The listening test results showed that the proposed small PLPCNet provided similar perceptual quality to a large generic LPCNet, achieving a 34\% reduction in model size. It also provided superior perceptual quality to the same-sized generic LPCNet. The smaller the size was, the more sound quality improvement PLPCNet achieved. In future work, we will expand the personalization concept to other NSC models and investigate its benefits in terms of bitrate reduction. PLPCNet was the first personalized neural speech codec proposed in the literature to our best knowledge.

\bibliographystyle{IEEEbib}
\bibliography{refs, mjkim}

\begin{thebibliography}{10}

\bibitem{OordA2016wavenet}
A.~{van den Oord}, S.~Dieleman, H.~Zen, K.~Simonyan, O.~Vinyals, A.~Graves,
  N.~Kalchbrenner, A.~Senior, and K.~Kavukcuoglu,
\newblock ``{WaveNet: A Generative Model for Raw Audio},''
\newblock in {\em Proc. 9th ISCA Workshop on Speech Synthesis Workshop (SSW
  9)}, 2016, p. 125.

\bibitem{KleijnW2018wavenet}
W.~B. Kleijn, F.~S.~C. Lim, A.~Luebs, J.~Skoglund, F.~Stimberg, Q.~Wang, and
  T.~C. Walters,
\newblock ``Wave{N}et based low rate speech coding,''
\newblock in {\em Proc. of the IEEE International Conference on Acoustics,
  Speech, and Signal Processing (ICASSP)}, 2018, pp. 676--680.

\bibitem{GarbaceaC2019vqvae}
Y.~Li C.~Garbacea, A.~{van den Oord},
\newblock ``Low bit-rate speech coding with {VQ-VAE} and a {WaveNet} decoder,''
\newblock in {\em Proc. of the IEEE International Conference on Acoustics,
  Speech, and Signal Processing (ICASSP)}, 2019.

\bibitem{ValinJ2019lpcnet}
J.-M. Valin and J.~Skoglund,
\newblock ``{LPCNet}: Improving neural speech synthesis through linear
  prediction,''
\newblock in {\em Proc. of the IEEE International Conference on Acoustics,
  Speech, and Signal Processing (ICASSP)}, 2019.

\bibitem{KalchbrennerN2018wavernn}
N.~Kalchbrenner, E.~Elsen, K.~Simonyan, S.~Noury, N.~Casagrande, E.~Lockhart,
  F.~Stimberg, A.~{van den Oord}, S.~Dieleman, and K.~Kavukcuoglu,
\newblock ``Efficient neural audio synthesis,''
\newblock in {\em Proc. of the International Conference on Machine Learning
  (ICML)}, 2018, vol.~80, pp. 2410--2419.

\bibitem{ValinJ2019lpcnetcoding}
J.-M. Valin and J.~Skoglund,
\newblock ``A real-time wideband neural vocoder at 1.6 kb/s using {LPCNet},''
\newblock in {\em Proc. Interspeech}, 2019.

\bibitem{Subramani2022e2elpcnet}
K.~Subramani, J.-M. Valin, U.~Isik, P.~Smaragdis, and A.~Krishnaswamy,
\newblock ``{End-to-End LPCNet}: A neural vocoder with fully-differentiable
  {LPC} estimation,''
\newblock {\em arXiv preprint arXiv:2202.11301}, 2022.

\bibitem{Valin2022lpcnetimproving}
J.-M. Valin, U.~Isik, P.~Smaragdis, and A.~Krishnaswamy,
\newblock ``Neural speech synthesis on a shoestring: Improving the efficiency
  of {LPCNet},''
\newblock in {\em Proc. of the IEEE International Conference on Acoustics,
  Speech, and Signal Processing (ICASSP)}, 2022.

\bibitem{Zeghidour2021soundstream}
N.~Zeghidour, A.~Luebs, A.~Omran, J.~Skoglund, and M.~Tagliasacchi,
\newblock ``Soundstream: An end-to-end neural audio codec,''
\newblock {\em IEEE/ACM Trans. Audio, Speech and Lang. Proc.}, vol. 30, pp.
  495–507, jan 2022.

\bibitem{defossez2022highfi}
A.~Défossez, J.~Copet, G.~Synnaeve, and Y.~Adi,
\newblock ``High fidelity neural audio compression,''
\newblock {\em arXiv preprint arXiv:2210.13438}, 2022.

\bibitem{VaswaniA2017transformer}
A.~Vaswani, N.~Shazeer, N.~Parmar, J.~Uszkoreit, L.~Jones, A.~N. Gomez,
  {\L}.~Kaiser, and I.~Polosukhin,
\newblock ``Attention is all you need,''
\newblock in {\em Advances in Neural Information Processing Systems (NIPS)},
  2017.

\bibitem{JiangX2023tf-codec}
X.~Jiang, X.~Peng, H.~Xue, Y.~Zhang, and Y.~Lu,
\newblock ``Latent-domain predictive neural speech coding,''
\newblock {\em IEEE/ACM Transactions on Audio, Speech, and Language
  Processing}, vol. 31, pp. 2111--2123, 2023.

\bibitem{kumar2023high}
R.~Kumar, P.~Seetharaman, A.~Luebs, I.~Kumar, and K.~Kumar,
\newblock ``High-fidelity audio compression with improved {RVQGAN},''
\newblock {\em arXiv preprint arXiv:2306.06546}, 2023.

\bibitem{KorseS2022postgan}
S.~Korse, N.~Pia, K.~Gupta, and G.~Fuchs,
\newblock ``{PostGAN}: A gan-based post-processor to enhance the quality of
  coded speech,''
\newblock in {\em Proc. of the IEEE International Conference on Acoustics,
  Speech, and Signal Processing (ICASSP)}, 2022, pp. 831--835.

\bibitem{SivaramanA2022jstsp}
A.~Sivaraman and M.~Kim,
\newblock ``{Efficient Personalized Speech Enhancement Through Self-Supervised
  Learning},''
\newblock {\em IEEE Journal of Selected Topics in Signal Processing}, vol. 16,
  no. 6, pp. 1342--1356, 2022.

\bibitem{SivaramanA2021waspaa}
A.~Sivaraman and M.~Kim,
\newblock ``Zero-shot personalized speech enhancement through speaker-informed
  model selection,''
\newblock in {\em Proc. of the IEEE Workshop on Applications of Signal
  Processing to Audio and Acoustics (WASPAA)}, 2021.

\bibitem{KimSW2021waspaa}
S.~Kim and M.~Kim,
\newblock ``Test-time adaptation toward personalized speech enhancement:
  Zero-shot learning with knowledge distillation,''
\newblock in {\em Proc. of the IEEE Workshop on Applications of Signal
  Processing to Audio and Acoustics (WASPAA)}, 2021.

\bibitem{Thakker2022arxiv}
M.~Thakker, S.~E. Eskimez, T.~Yoshioka, and H.~Wang,
\newblock ``Fast real-time personalized speech enhancement: End-to-end
  enhancement network ({E3Net}) and knowledge distillation,''
\newblock {\em arXiv preprint arXiv:2204.00771}, 2022.

\bibitem{usac1}
{{ISO/IEC DIS} 23003-3},
\newblock ``Information technology -- {MPEG} audio technologies -- part 3:
  Unified speech and audio coding,'' 2011.

\bibitem{evs}
{ETSI TS 126 445 V13. 2.0},
\newblock ``{Universal Mobile Telecommunications System (UMTS); LTE; Codec for
  Enhanced Voice Services (EVS); Detailed algorithmic description (3GPP TS
  26.445 version 13.2. 0 Release 13},'' 2016.

\bibitem{PanayotovV2015Librispeech}
V.~Panayotov, G.~Chen, D.~Povey, and S.~Khudanpur,
\newblock ``{Librispeech}: {An} {ASR} corpus based on public domain audio
  books,''
\newblock in {\em Proc. of the IEEE International Conference on Acoustics,
  Speech, and Signal Processing (ICASSP)}, 2015, pp. 5206--5210.

\bibitem{BromleyJ1994nips}
J.~Bromley, I.~Guyon, Y.~Le{C}un, E.~S{\"a}ckinger, and R.~Shah,
\newblock ``Signature verification using a ``siamese" time delay neural
  network,''
\newblock in {\em Advances in Neural Information Processing Systems (NIPS)},
  1994, pp. 737--744.

\bibitem{ChiccoD2021siamese}
D.~Chicco,
\newblock {\em {Siamese Neural Networks: An Overview}}, pp. 73--94,
\newblock Springer US, New York, NY, 2021.

\bibitem{VanDerMaatenL2008tsne}
L.~Van der Maaten and G.~Hinton,
\newblock ``{Visualizing Data using t-SNE},''
\newblock {\em Journal of Machine Learning Research}, vol. 9, no. 11, 2008.

\bibitem{KingmaD2015adam}
D.P. Kingma and J.~Ba,
\newblock ``Adam: A method for stochastic optimization,''
\newblock in {\em Proc. of the International Conference on Learning
  Representations (ICLR)}, 2015.

\bibitem{mushra}
{{ITU-R} {Recommendation} {BS} 1534-3},
\newblock ``Method for the subjective assessment of intermediate quality levels
  of coding systems ({MUSHRA}),'' 2015.

\end{thebibliography}

\end{document}